\documentclass[aps,prb,twocolumn,groupedaddress,showpacs]{revtex4}
\usepackage{epsfig}
\usepackage{color}

\begin{document}

\title{Pair formation and collapse in imbalanced Fermion populations with unequal masses}

\author{G. G. Batrouni$^1$, M. J. Wolak$^2$,
  F. H\'ebert$^1$, V. G. Rousseau$^3$}

\affiliation{$^1$INLN, Universit\'e de Nice-Sophia Antipolis, CNRS; 
1361 route des Lucioles, 06560 Valbonne, France}

\affiliation{$^2$Centre for Quantum Technologies, National University
  of Singapore, Singapore 117542}

\affiliation{$^3$Lorentz-Instituut, Universiteit Leiden, Postbus 9504,
  2300 RA, Leiden, The Netherlands}

\begin{abstract}
We present an exact Quantum Monte Carlo study of the effect of unequal
masses on pair formation in Fermionic systems with population
imbalance loaded into optical lattices. We have considered three forms
of the attractive interaction and find in all cases that the system is
unstable and collapses as the mass difference increases and that the
ground state becomes an inhomogeneous collapsed state. We also address
the question of canonical {\it vs} grand canonical ensemble and its
role, if any, in stabilizing certain phases.

\end{abstract}


\maketitle

\section{Introduction}

The properties of fermionic systems with imbalanced populations have
long been of interest for a variety of reasons. Of particular
importance is the pairing mechanism resulting in composite bosonic
particles and leading to superconductivity/superfluidity. The original
focus~\cite{fulde64,larkin64} was on polarized superconductors with an
imbalance between the two spin populations where it was predicted that
the Cooper pairs form with a center of mass momentum equal to the
difference of the Fermi momenta of the two populations. This
Fulde-Ferrell-Larkin-Ovchinnikov (FFLO) phase \cite{fulde64,larkin64}
proved to be difficult to observe in solids and was only achieved
recently in heavy fermion systems \cite{radovan03}. 

The recent experimental realization of trapped ultra-cold fermionic
atoms with tunable attractive interactions has intensified interest in
this problem. Experiments, in which two hyperfine states of ultra-cold
fermionic atoms play the role of up and down spins, have demonstrated
the presence of pairing in the case of unequal
populations.\cite{zwierlein06,partridge06} This has spurred much
theoretical work using a variety of approaches such as mean field
\cite{castorina05,sheehy06,kinnunen06,machida06,gubbels06,parish07,hu07,he07,wilczek,koponen},
effective Lagrangian,\cite{son06} bosonisation \cite{boso} and Bethe
ansatz \cite{orso07} applied to the uniform system with extensions to
the trapped system using the local density approximation (LDA). 

For the uniform case, the controversy continues on the details of the
paired state: FFLO pairs forming with non-zero momentum {\it vs}
breached pairing (BP) at zero momentum\cite{koponen,wilczek,sarma} and
on the robustness of such phases. Distinguishing features between
these two possibilities include the presence, for FFLO, of spatial
modulations (inhomogeneity) in the pairing order parameter, and
consequently a peak at non-zero momentum in the pair momentum
distribution function as opposed to the coexistence of a superfluid
and normal component in a translationally invariant and isotropic
state in the BP scenario.  In addition, excitations are gapped for the
FFLO phase but not so for BP.

No general consensus has emerged on which pairing type dominates.
However, recent exact numerical work on the ground state of
one-dimensional Fermi systems with imbalanced populations loaded into
optical lattices has found that FFLO is the only pairing mechanism
both in the uniform~\cite{batrouni,luscher,rizzi} and the
trapped~\cite{batrouni,luscher,tezuka07,feiguin07} systems. Similar
results have been found without the optical lattice.\cite{casula} On
the other hand, approximate dynamical mean field (DMFT) results appear
to indicate that no FFLO phase is present in the three-dimensional
system.\cite{dao}

In addition to the above equal-mass cases, of relevance in condensed
matter and trapped ultra-cold atomic systems, there is great interest
in the case of unequal masses. This arises naturally in trapped
mixtures of different atomic species, {\it e.g.} K-Rb, and also in
cold dense quark matter where the $c,b$ and $t$ quarks are much
heavier than the $u,d$ and $s$ quarks. It was argued theoretically
that such mass and population imbalance leads to the BP
phase~\cite{wilczek1,wilczek2} with gapless excitations which can be
stabilized by longer-range interactions.\cite{wilczek3} In this paper
we examine this question with exact Quantum Monte Carlo (QMC)
simulations. We focus on the one-dimensional system loaded into an
optical lattice governed by the extended Hubbard Hamiltonian

\begin{eqnarray}
  \label{Hamiltonian}
\hat\mathcal H=&-&\sum_{l \sigma}t_{\sigma} (c_{l+1\,\sigma}^{\dagger}
c_{l\,\sigma}^{\phantom\dagger} + c_{l\,\sigma}^\dagger c_{l+1 \,
  \sigma}^{\phantom\dagger}) \nonumber \\ &+&
\sum_{i,j,\sigma,\sigma^{\prime}} U_{ij,\sigma
  \sigma^{\prime}}n_{i\,\sigma} n_{j\,\sigma^{\prime}},
\end{eqnarray}
where $c_{j\,\sigma}^\dagger(c_{j\, \sigma})$ are fermion creation
(annihilation) operators on spatial site $j$ with the fermionic
species labeled by $\sigma=1,2$ and
$n_{j\,\sigma}=c_{j\,\sigma}^\dagger c_{j\, \sigma}$ is the
corresponding number operator. The unequal masses are embodied in the
unequal hopping parameters. We set the energy scale by taking $t_1=1$
and studying the system as $t_2$ gets smaller ($m_2$ gets larger). In
general, the interaction term $U_{ij,\sigma \sigma^{\prime}}$ can
couple all fermions on all sites; in what follows we shall consider
three different forms.

For our simulations, we use a continuous imaginary time canonical
``worm" (CW) algorithm where the total number of particles is
maintained strictly constant.\cite{pollet} In this algorithm, two
worms are propagated, one for each type of fermion, which allows the
calculation of the real-space Green functions of the two fermionic
species, $G_\sigma$, and the pair Green function, $G_{\rm pair}$,
\begin{eqnarray}
G_\sigma(l) &=& \langle c_{j+l \,\sigma}^{\phantom\dagger} c_{j
\,\sigma}^\dagger \rangle, \nonumber \\ G_{\rm pair}(l) &=& \langle
\Delta_{j+l}^{\phantom\dagger}\,\Delta^{\dagger}_{j} \rangle,
\nonumber \\ \Delta_j &=& c_{j \, 2} \,c_{j \, 1}.
\end{eqnarray}
It is then immediate to obtain the momentum distributions
$n_{\sigma}(k)$ and $n_{\rm pair}(k)$ from their Fourier
transforms. We emphasize that this algorithm is exact: There are no
approximations and the only errors are statistical (which typically
correspond to the size of the symbols).  The polarization is given by
$P=(N_2-N_1)/(N_2+N_1)$, where $N_1(N_2)$ is the minority (majority)
particle numbers.  The associated Fermi wave vectors are $k_{\rm F
\sigma} = \pi N_\sigma/L$. We ran our simulations at $\beta=2L$, where
$L$ (typically $L=32$) is the number of lattice sites. We have
verified that these values are large enough to yield the ground state
properties, which is our focus. Typical runs takes a day or two on a
desktop computer.

\begin{figure}[t]
\centerline{\epsfig{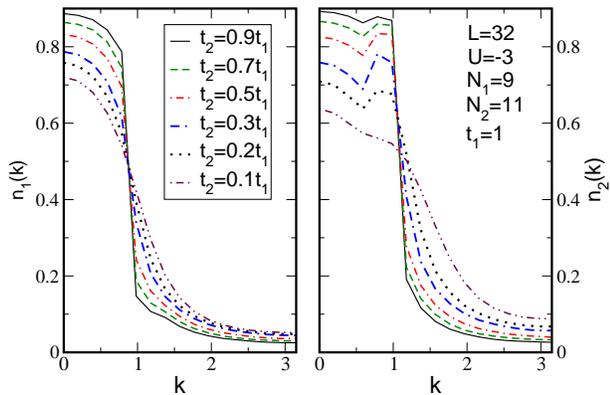}}
\caption{(color online).  The momentum distribution for the minority
(left) and majority (right) populations.}
\label{fig1}
\end{figure}

\section{Heavy Majority: $t_1>t_2$}

\begin{figure}[t]
\centerline{\epsfig{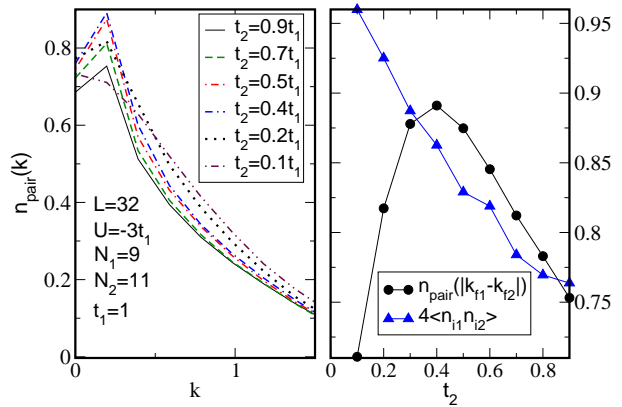}}
\caption{(color online). Left: The pair momentum distribution
  function. Right: The circles show the height of $n_{\rm
    pair}(|k_{F_1}-k_{F_2}|)$ vs $t_2$, the triangles show $\langle
  n_{i1} n_{i2}\rangle$ scaled by 4 for better visibility. }
\label{fig2}
\end{figure}

We begin with the simplest form for the interaction term,
$U_{ij,\sigma
\sigma^{\prime}}=U\delta_{i,j}(1-\delta_{\sigma,\sigma^{\prime}})$,
{\it i.e.} only contact attraction ($U<0$) between the two
species. Figure~\ref{fig1} shows what happens to the minority (left)
and majority (right) momentum distributions as $t_2$ is decreased for
fixed $U=-3t_1$ and $N_1=9,\, N_2=11$. For $t_2=0.9 t_1$,
$n_{\sigma}(k)$ drops sharply at the respective Fermi momenta,
$(k_{F_1},\,k_{F_2})$ while at the same time, the pair momentum
distribution, $n_{\rm pair}(k)$, exhibits a maximum at
$k=|k_{F_1}-k_{F_2}|$ (fig.~\ref{fig2} left). This indicates pair
formation at non-zero center-of-mass momentum and thus an FFLO
phase.\cite{batrouni,luscher,rizzi,tezuka07,feiguin07} As the mass of
the majority particles increases, $t_2$ decreases, the attractive
interaction effectively increases because $|U|/t_2$ increases. The
increased binding is demonstrated by $\langle n_{i1} n_{i2}\rangle$
which takes values between $N_1N_2/L^2$ (no binding) and $N_1/L$ (all
minority particles have formed pairs)~\cite{batrouni}. This quantity
is shown in fig.~\ref{fig2} (triangles, right panel) scaled by $4$ for
better visibility. Clearly, as $t_2$ decreases, $\langle n_{i1}
n_{i2}\rangle$ increases toward its upper limit. As a consequence of
this increased binding, the FFLO effect first intensifies, its peak at
$k=|k_{F_1}-k_{F_2}|$ increases in height, reaches a maximum for
$t_2=0.4t_1$, then decreases and disappears as shown in
fig.~\ref{fig2} (circles, right panel).  Note, for example, that for
$t_2=0.1t_1$ the peak of $n_{\rm pair}(k)$, fig.~\ref{fig2} (left), is
at $k=0$.

Also noteworthy is the appearance of a dip in $n_2(k)$, at $k\approx
k_{F_1}$, whose depth increases with decreasing $t_2$, reaching a
maximum for $t_2=0.4t_1$, corresponding to the maximum FFLO
effect. Further decreasing $t_2$ washes this out. This feature can be
understood as follows. As $t_2$ decreases, the interaction between the
minority and majority effectively increases ($|U|/t_2$), thus
increasing the depletion of $n_{\sigma}(k<k_{F_{\sigma}})$. Most of
the depletion for both $n_1(k)$ and $n_2(k)$ happens for $k<k_{F_1}$
since this is where most of the minority resides. This depletion for
$k<k_{F_1}$ leaves $n_2(k)$ with a bump for $k_{F_1}<k<k_{F_2}$.

\begin{figure}[t]
\centerline{\epsfig{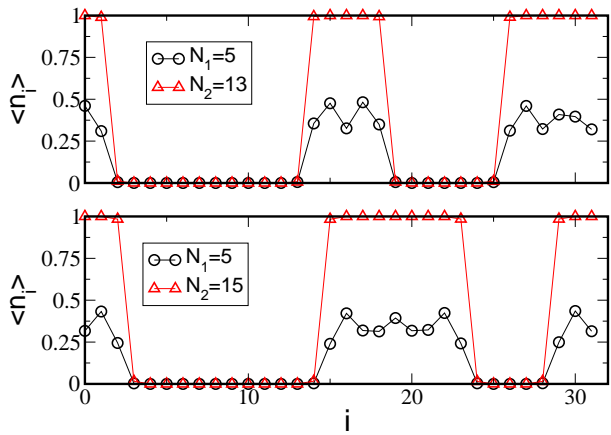}}
\caption{(color online). Average density profiles for collapsed
  systems. Here $U=-7t_1$ and $t_2=0.04t_1$. Note the Friedel-like
  oscillations in $n_{i1}$.}
\label{fig3}
\end{figure}

At first glance, it might seem that the disappearance of the FFLO
peak, $n_{\rm pair}(k=|k_{F_1}-k_{F_2}|)$, as $t_2$ decreases and the
appearance of a peak $n_{\rm pair}(k=0)$ signals pair formation at
$k=0$ and thus a BP phase. However, a simple argument offers another
option. The Fermi momentum of the majority is $k_{F_2}=\pi N_2/L$;
consequently, as these particles get heavier, $t_2$ smaller, their
kinetic energy becomes negligible and can be ignored. To minimize its
free energy, the system will then optimize the potential and kinetic
energies of the light particles. The optimal potential energy is
obtained when the light particles are on the same sites as the heavy
ones. On the other hand, the kinetic energy is optimized when the
light particles are delocalized. Both these energies can be optimized
if the heavy particles coalesce, forming a contiguous region with one
heavy particle per site, $n_{i2}=1$. This region, then, acts as a
platform on which the light particles can delocalize over its entire
extent while always being in contact with the heavy particles thus
minimizing their potential energy. This is indeed what happens as
the density profiles, $n_{i1}$ and $n_{i2}$, show in fig.~\ref{fig3}
for two polarizations. To summarize, as $t_2$ decreases, the system
goes from an FFLO phase to a spatially collapsed one.

\begin{figure}[t]
\centerline{\epsfig{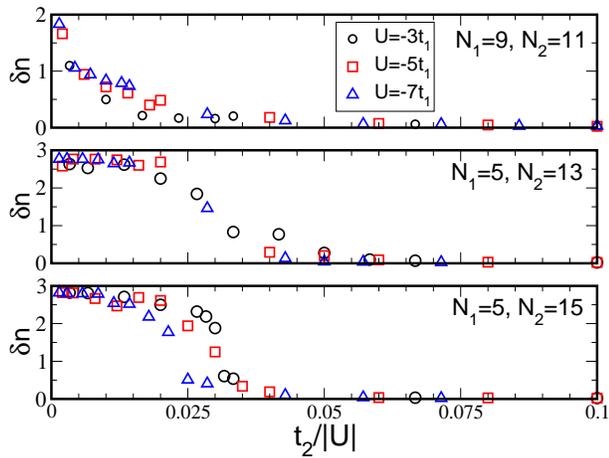}}
\caption{(color online). $\delta n$ versus $t_2/|U|$ showing the
  collapse as $t_2$ gets small enough.}
\label{fig4}
\end{figure}

To quantify this collapse, we define the quantity~\cite{foot1}
$(\delta n)^2\equiv \sum_i(\langle n_{i1} \rangle - N_1/L )^2$ which
is essentially zero when the system is uniform ($\langle
n_{i1}\rangle=N_1/L$) and grows as the collapse happens. In
fig.~\ref{fig4} we show $\delta n$ versus $t_2/|U|$ for three
polarizations and three couplings in each case. We see that, indeed,
the system collapses as $t_2$ decreases and that for a given
polarization, this collapse appears to happen at approximately the
same value of $t_2/|U|$. Also, the larger the polarization, the easier
it is to trigger the collapse (larger $t_2$). This behavior holds for
all the parameters we examined, specifically $-13<U\leq -1$, and
polarizations $0.1\leq P \leq 0.55$

We, therefore, conclude that in the presence of only contact
attraction, the BP phase is {\it not} realized as the majority
population is made heavier; instead the system collapses. To counter
the tendency of the heavy particles to clump together as in
fig.~\ref{fig3}, we introduce longer-range interaction. It is
reasonable to suppose that near neighbor (nn) repulsion between
particles of the same species would tend to oppose such collapse. To
this end we redid the above study but with the interaction term
$H_I=\sum_i Un_{i1}n_{i2}+V\sum_{i,\sigma}n_{i\sigma}n_{i+1\sigma}$
with $U<0$ and $V>0$. We studied this for $0.1 \leq P\leq 0.44$ for
$-10t_1\leq U\leq -4t_1$ and found that, indeed, the stability of the
system against collapse is extended to smaller values of $t_2$ but
that eventually the system always collapses. Furthermore, before the
collapse, the system always exhibits FFLO pairing while after
collapse, the nn repulsion term, $V$, produces density oscillations as
the presence of near neighbors is opposed. Therefore, this second form
of the interaction also fails to produce the BP phase.

\begin{figure}[t]
\centerline{\epsfig{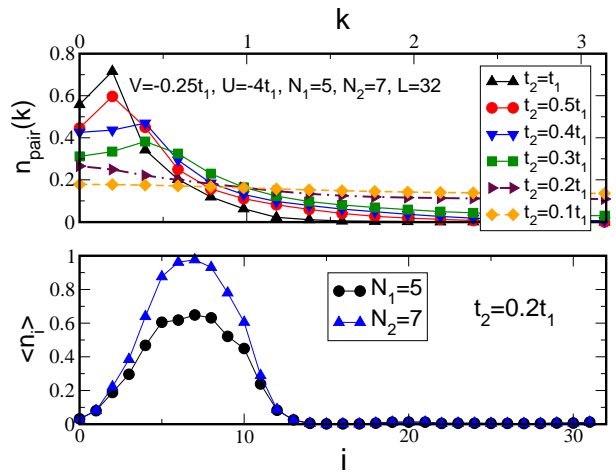}}
\caption{(color online). Top: The pair momentum distribution for
  several values of $t_2$ showing the disappearance of the FFLO
  peak. Bottom: Average density profile for same parameters as top
  panel but $t_2=0.2t_1$ showing the system when it first
  collapses. The contact interaction is $U=-4t_1$ and the nn
  interaction is $V=-0.25t$ acting only between particles of opposite
  spin.}
\label{fig5}
\end{figure}

\begin{figure}[h]
\centerline{\epsfig{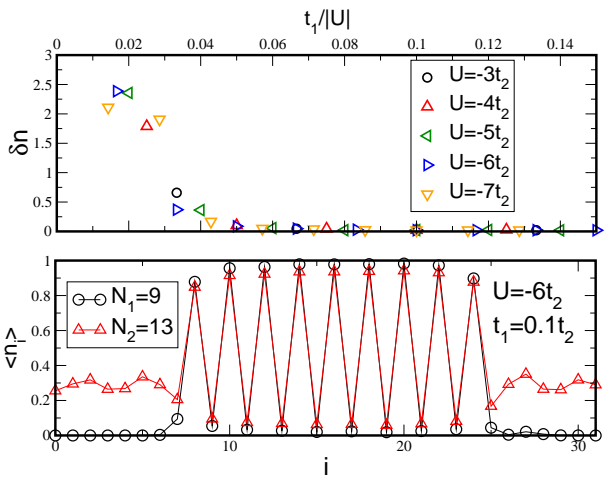}}
\caption{(color online). Top: $\delta n$ rises sharply at 
$t_1/|U|\approx 0.03$ signalling a transition to an
inhomogeneous density profile. Bottom: For $t_1/|U|<0.03$,
the system undergoes phase separation as shown in this
typical density profile. Part of the system is in a charge
density wave phase while the remainder is in a free fermion
phase.
}
\label{fig6}
\end{figure}

\begin{figure}[h]
\centerline{\epsfig{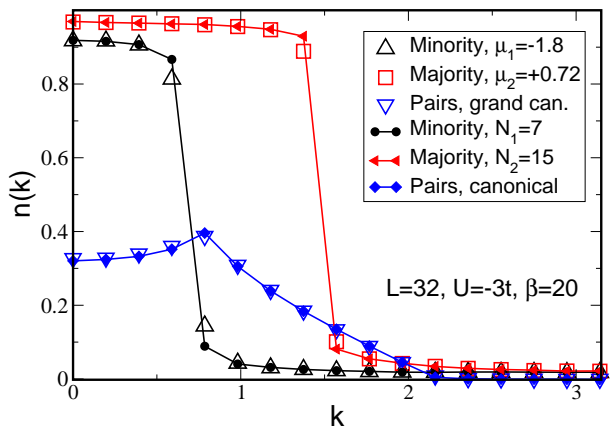}}
\caption{(color online). Momentum distributions for the minority,
  majority and pair populations using canonical and grand canonical
  QMC with $t_1=t_2$. In the grand canonical case, $\langle
  N_1\rangle=6.996\, (\mu_1=-1.8)$, $\langle N_2\rangle = 14.949\,
  (\mu_2=+0.72)$. }
\label{fig7}
\end{figure}

We now turn to the third interaction form which was proposed in
Ref.~\cite{wilczek3} and argued to yield the BP phase. This
form extends beyond the contact term and acts only between particles
from different species. The idea is based on the assumption of three
competing homogeneous phases: a normal state of free fermions, a fully
gapped BCS superfluid and a gapless BP phase.  It was argued that by
giving the interaction term structure in momentum space, one may be
able to stabilize the BP phase; a Gaussian dependence on distance was
used for the interaction potential~\cite{wilczek3}.  In our
simulations we cut the range at the nn level and, therefore, have
$H_I=\sum_i (Un_{i1}n_{i2}+Vn_{i1}n_{i+1 2})$. Note that, unlike the
previous form, $V$ here acts only between $n_1$ and $n_2$. In momentum
space, this has the form $U+2V{\rm cos}(q)$ at momentum $q$ and, with
$U<0$ and $V<0$, is more attractive at small $q$ in contradistinction
to the pure contact interaction which is independent of $q$. Our exact
QMC results show, however, that the system is still unstable and
collapses as $t_2$ decreases, just like in the contact interaction
case. This is shown in fig.~\ref{fig5}: No BP phase has been found. In
fact, with the longer-range attraction, the system is even {\it more}
unstable as can be easily understood by the same real space argument
presented above for the contact interaction case.

\section{Heavy Minority: $t_1 < t_2$}

So far, we have considered the case where the heavy particles are the
majority which leads to collapsed configurations like those in
Fig.~\ref{fig3}. The question then is: Will the system still collapse
when the heavy particles are the minority and what form will the
collapsed configurations take?  We now consider this case with a heavy
minority population $N_1=9$, and a lighter majority, $N_2=13$, on a
$32$-site lattice with $\beta=64$. As before, we fix the light
population hopping $t_2=1$ to define the energy scale and we study the
collapse as a function of the heavy minority hopping parameter,
$t_1/t_2$, and the attractive interaction, $U/t_2<0$. In the top panel
of fig.~\ref{fig6} we show, like in fig.~\ref{fig4}, $\delta n$ as a
function $t_1/|U|$ for five different values of the interaction,
$U$. The behavior is similar to that in fig.~\ref{fig4}; we find that
for these values of $N_1$ and $N_2$, $\delta n$ increases sharply for
$t_1/|U|\approx 0.03$ signalling spatial collapse in the system. One
candidate for the collapsed confiugration is that, as before, the
heavy particles (the minority in this case) form a contiguous region
thus providing a platform for the lighter particles. This would then
result in a contiguous region with one heavy and one light particle
per site and the excess light majority particles spread over the rest
of the system. This, however, does not minimize the energy because the
light particles residing on the heavy particle platform are blocked:
They are in a Mott state and have zero kinetic energy. In the bottom
panel of fig.~\ref{fig6} we show a typical density profile of a
collapsed configuration. It is easy to understand energetically why
this density wave structure is favored over the previous candidate:
The light particles are never blocked in a Mott region and can always
hop to neighboring sites to optimize the free energy.

We note that the configuration in fig.~\ref{fig6} corresponds to phase
separation: In one region the system is in a charge density wave phase,
in the other region it has free fermions. These two phases co-exist.
On the other hand, the configurations in fig.~\ref{fig3} correpsond
to spatial collapse: The system has a regions void of particles and
regions where all the particles reside.

\section{Canonical vs Grand Canonical}

Finally, we consider the question of canonical versus grand canonical
ensembles. It has been suggested~\cite{wilczek3,koponen} that the
stability of BCS, FFLO or BP phases can depend on whether one fixes
the populations or the chemical potentials. All the above results have
been obtained with the CW algorithm where $N_1$ and $N_2$ are kept
strictly fixed. An alternative is to add the chemical potential term,
$\sum_i(-\mu_{1}n_{i1}-\mu_2 n_{i2})$, to the Hamiltonian,
eq.(\ref{Hamiltonian}), and use a grand canonical QMC algorithm such
as the Determinant Quantum Monte Carlo (DQMC)
algorithm.\cite{blankenbecler81} We compare in fig.~\ref{fig7} the
momentum distributions obtained with the CW and the DQMC algorithms
with $\mu_1$ and $\mu_2$ chosen to give average fillings corresponding
to the fixed fillings in the canonical case. There is no disagreement
between the two; in particular, the FFLO phase is seen to be stable
whether one fixes the populations or the chemical potentials. Of
course, if one changes $t_2$ while holding the chemical potentials
fixed, the average populations, $\langle N_1\rangle$ and $\langle
N_2\rangle$, will change. Nonetheless, whatever $\langle N_1\rangle$
and $\langle N_2\rangle$ one has obtained by fixing $\mu_1$ and
$\mu_2$, one will obtain the same physics in the canonical ensemble by
fixing the populations to the corresponding values, as seen in
fig.~\ref{fig7}. One is free to study these phases and their stability
in either ensemble.

\section{Conclusions}

In summary, we have studied, using exact QMC simulations, the effect
of mass differences between two imbalanced Fermion populations. For
the case where the majority population is the heavier, we performed
our study with three possible attractive interaction terms. In all
three cases, we have found the system to be unstable and to collapse
when the mass disparity is large enough: The BP phase is not realized
by tuning the mass ratio between the two populations. For the case
where the minority is heavier, we showed the system still collapses
when the mass difference is large enough, but in this case it forms
density wave structures. We have also shown that fixing the
populations or the chemical potentials leads to the same physics. The
stabilization of sought-after phases is not favored by one ensemble
rather than another.

\acknowledgments
This work is supported by: the CNRS (France) PICS 3659 and PICS 4159,
a PHC Merlion grant (SpinCold 2.02.07), the National Research
Foundation and the Ministry of Eduction (Singapore) and by the
research program of the `Stichting voor Fundamenteel Onderzoek der
Materie (FOM)'. We acknowledge very helpful discussions with
K. Bouadim and R. T. Scalettar, C. Miniatura and B. G. Englert.


\begin{thebibliography}{0}
\bibitem{foot1} One can also use $n_2(x)$, but the behavior is
  identical. We only show $\delta n$ defined with $n_1(x)$.
\end{thebibliography}
\end{document}